\documentclass{moriond}

\def\be{\begin{equation}}
\def\ee{\end{equation}}
\def\bea{\begin{eqnarray}}
\def\eea{\end{eqnarray}}

\usepackage{amsmath}
\usepackage{xspace}
\usepackage{caption}
\usepackage{subcaption}

\graphicspath{{figures/}}

\newcommand{\nlzero}{\ensuremath{N_L^{(0)}}\xspace}
\newcommand{\rdnlzero}{\ensuremath{\mathrm{RD{\text -}}N_L^{(0)}}\xspace}

\newcommand{\nlone}{\ensuremath{N_L^{(1)}}\xspace}

\newcommand{\cpp}[0]{\ensuremath{C^{\phi\phi}_L}\xspace}

\newcommand{\lcdm}{\ensuremath{\mathrm{\Lambda CDM}}\xspace}

\newcommand{\Photo}{}

\begin{document}
\vspace*{4cm}
\title{CMB LENSING POWER SPECTRUM WITH NEXT GENERATION SURVEYS}

\author{L. LEGRAND and J. CARRON}

\address{Universit\'e de Gen\`eve, D\'epartement de Physique Th\'eorique et CAP, 24 Quai Ansermet, CH-1211 Gen\`eve 4, Switzerland}

\maketitle\abstracts{
       We introduce a new estimator of the CMB lensing power spectrum, together with its likelihood, based on iterative lensing reconstruction. Despite the increased complexity of the lensing maps, this estimator shares similarities with the standard quadratic estimator. Most importantly, it is unbiased towards the assumptions done on the noise and cosmology for the lensing reconstruction. This new spectrum estimator can double the constraints on the lensing amplitude compared to the quadratic estimator, while keeping numerical cost under control and being robust to errors.}

\section{Iterative lensing spectrum estimator}

Gravitational lensing of the CMB is a powerful probe of the growth of structures, and is expected to give tight constraints on the sum of neutrino masses. Current CMB experiments mostly rely on the well-known quadratic estimator (QE) to estimate the lensing potential. Next generation surveys, such as CMB-S4, will rely on a more efficient approach.
Indeed, since the primordial B-mode signal is small, and if the foreground and noise levels are well below the lensing B-mode power of $\sim 5\mu \textrm{K}$-arcmin, one could reconstruct perfectly the lensing field from the observed polarisation maps.
This can be achieved with a likelihood-based reconstruction\cite{Hirata:2003ka}. An implementation of the maximum a posteriori (MAP) lensing reconstruction using an iterative delensing procedure was developed in \cite{Carron:2017mqf}. In \cite{Legrand:2021qdu} we introduced a lensing spectrum estimator and its likelihood based on this MAP reconstruction. We summarize below our main results. 

The lensing spectrum estimated with a QE \cite{Hu:2001kj} is a four-point function of the CMB maps. It contains the lensing spectrum \cpp we want to measure, but also `bias' terms, dominated by \nlzero and \nlone which can be characterized analytically.
This imply that one can debias this QE spectrum to get an estimate of the true lensing spectrum. In practice a standard cosmology analysis \cite{Planck:2018lbu} uses a realisation dependant debiaser \rdnlzero, robust to the assumptions made on the experimental noise and fiducial cosmology.

\begin{figure}
       \centering
       \begin{subfigure}{0.5\textwidth}
         \centering
         \includegraphics[width=0.9\linewidth]{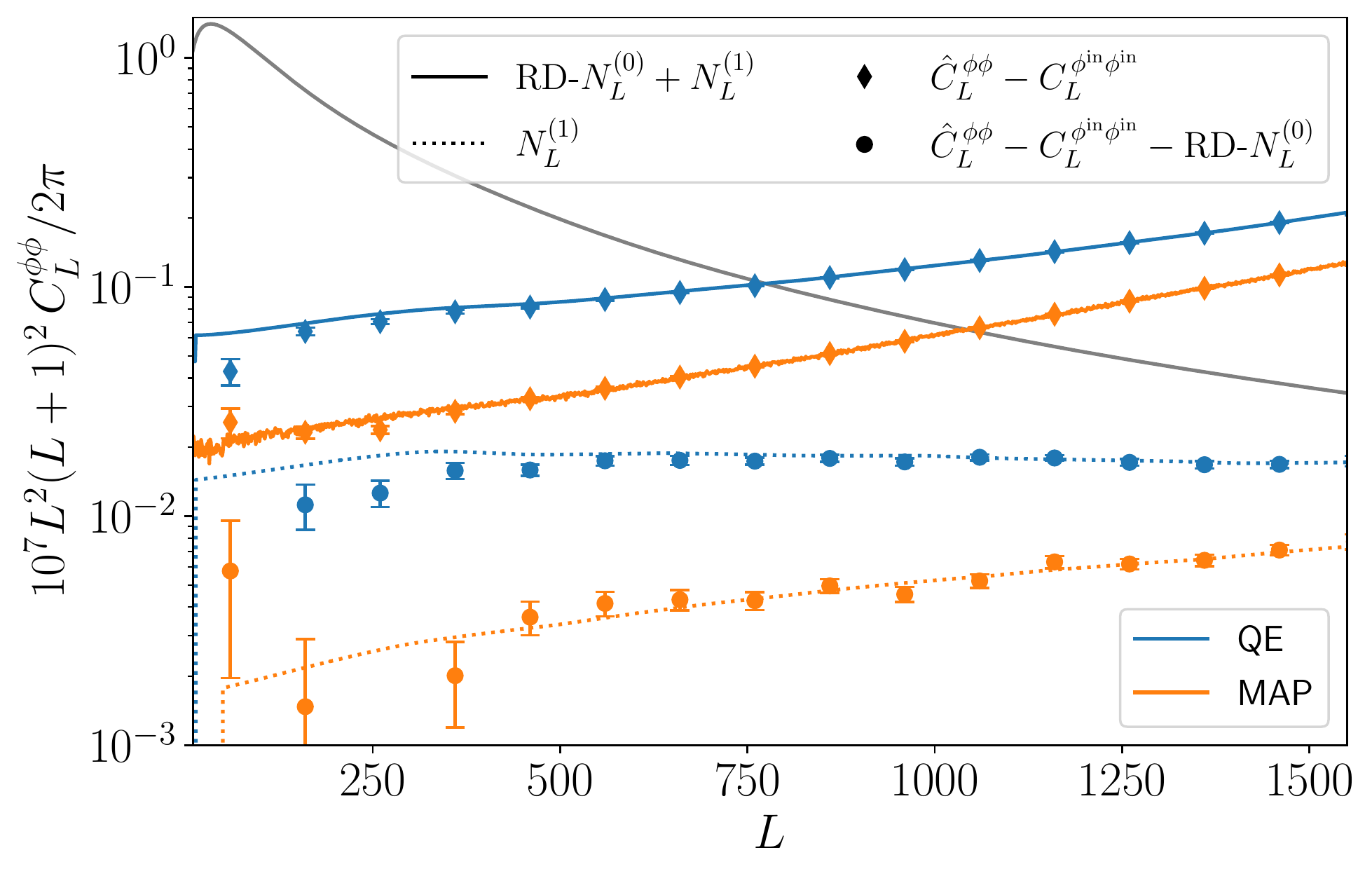}
       \end{subfigure}%
       \begin{subfigure}{.4\textwidth}
         \centering
         \includegraphics[width=\linewidth]{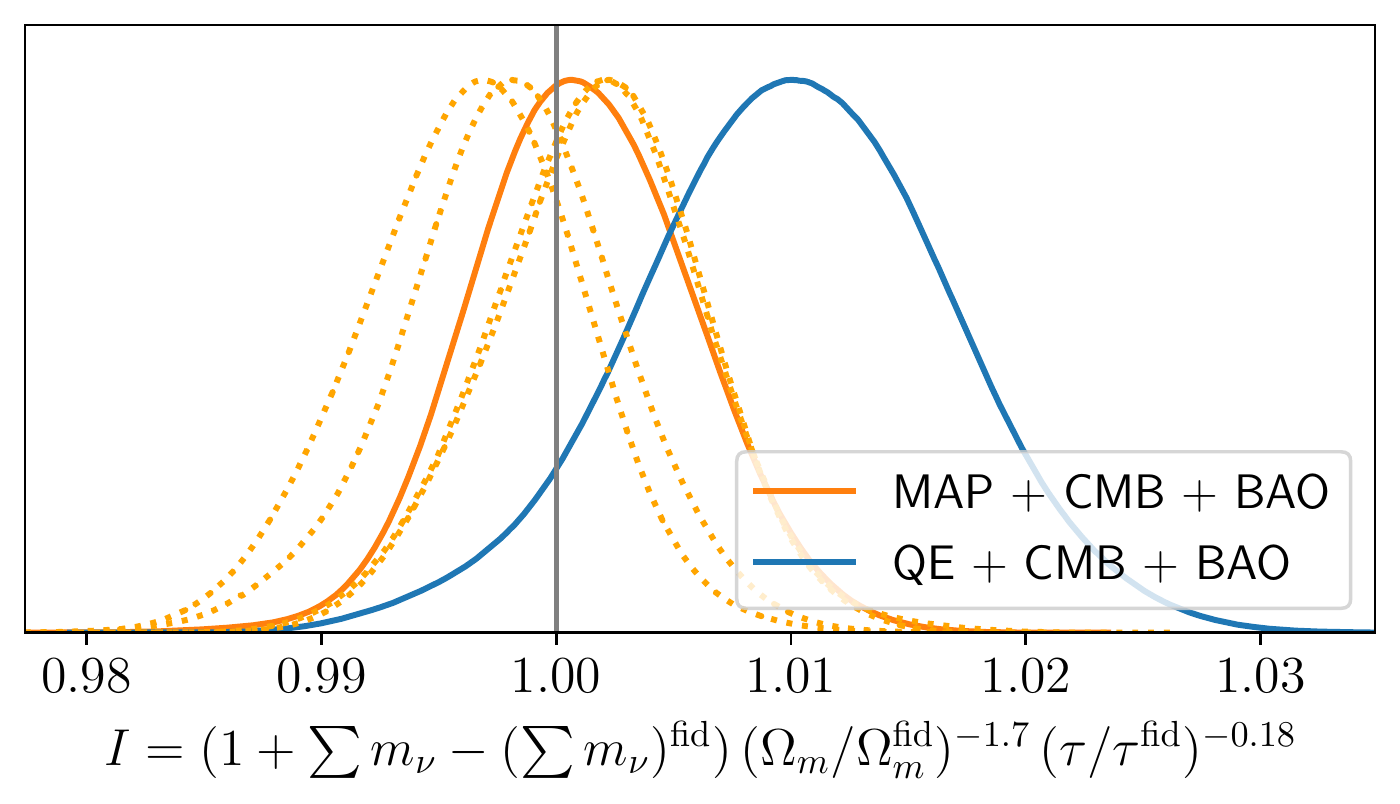}
       \end{subfigure}
       \caption{
              \textit{Left panel:} Diamonds show the estimated lensing spectra subtracted by the input spectrum (suppressing cosmic variance) for the QE and MAP estimators (resp. blue and orange), these should correspond to the sum of the \rdnlzero and \nlone biases, shown as solid lines. Circles show the estimated spectra subtracted by the input spectrum and by \rdnlzero, which should be dominated by the \nlone bias shown as the dotted lines. The grey line shows the fiducial lensing spectrum.
              \textit{Right panel:} Marginalized constraints on the derived parameter $I$ for a CMB-S4 experiment combined with DESI-BAO. Constraint using a QE is shown in blue, while constraints with the MAP estimator are shown in orange. 
              Dashed lines are four independent CMB realisations. 
              }
       \label{fig}
\end{figure}

The lensing potential field reconstructed with an iterative MAP estimator\cite{Carron:2017mqf} is a complex function of the CMB maps, and it is out of reach to track analytically the bias terms of its spectrum. 
In \cite{Legrand:2021qdu} we derive the \nlzero and \nlone bias terms of the MAP lensing spectrum using the same expressions as for the QE, but replacing the fiducial CMB and lensing spectra with partially delensed spectra. These delensed spectra are obtained in an iterative procedure until convergence.
We also introduce a \rdnlzero debiaser of the MAP estimate.
Left panel of Fig.\ref{fig} shows that we accurately estimate the bias terms of our MAP lensing spectrum.
We see also that the MAP \nlzero bias is lowered by a factor 2 compared to the QE, improving by the same amount the signal to noise ratio of the lensing spectrum amplitude. 

\section{Lensing likelihood}

We generate two datasets, each with a different cosmology, one being the fiducial cosmology used for the lensing reconstruction. 
We simulate a full-sky CMB-S4, and compute the QE and the MAP lensing fields. 
Our data-vectors are the pseudo spectrum debiased by \rdnlzero and \nlone. We assume they follow a Gaussian likelihood, and we estimate their covariances from 1024 flat-sky simulations, rescaled to get a $40\%$ sky fraction. 
We include the unlensed CMB and the DESI-BAO likelihoods, assuming all three likelihoods are independent.
We sample the six standard \lcdm parameters plus the sum of neutrino mass with a MCMC. For both datasets our MAP lensing spectrum likelihood is able to recover unbiased parameters estimates.
It appears that the marginalised constraints on $\sum m_\nu$ with a MAP does not improve compared to a QE. This could be due to remaining degeneracies between parameters. We perform a principal component analysis of our chains on the parameters $\sum m_\nu, \Omega_{\rm m}$ and $\tau$. We found that the combination  shown in the right panel of Fig.\ref{fig} gets a factor almost two reduction of the $1\sigma$ uncertainty from the QE to the MAP, reaching the statistical power of our new estimator. 

To conclude, we introduced a new CMB lensing power spectrum estimator, robust towards the ingredients assumed for the lensing reconstruction. This opens the door to improved and unbiased constraints on key cosmological parameters, such as the sum of neutrino mass.

\section*{Acknowledgments}

The authors acknowledge support from a SNSF Eccellenza Professorial Fellowship (No. 186879).

\end{document}